\documentclass[twocolumn,showpacs,amsmath,amssymb,superscriptaddress]{revtex4}
\usepackage{amsmath,amssymb,color,latexsym,natbib,graphicx}
\usepackage{dcolumn}
\usepackage{subfigure}
\usepackage{bm}
\usepackage{color}
\usepackage{ulem}

\newcommand{\be}{\begin{equation}}
\newcommand{\ee}{\end{equation}}
\def\ba{\begin{eqnarray}}
\def\ea{\end{eqnarray}}
\def\eg{{\it e.g.}\ }

\def\ie{{\it i.e.}\ }

\def \pmbmath{\mathpalette\pmbmathaux}
\def \pmbmathaux#1#2{
         \pmbtext{$#1#2$}}
\def \pmbtext#1{\leavevmode
     \setbox0\hbox{#1}
     \kern-0,2pt \copy0 \kern-\wd0
     \kern0,4pt \copy0 \kern-\wd0
     \kern-0,2pt \raise0,3pt \box0 }

\def\bell{{\pmbmath{\ell}}}

\def\kpe{k_{\perp}}
\def\kpeb{{\pmbmath{\kpe}}}
\def\kpa{k_{\parallel}}

\begin{document}
\renewcommand{\thesubfigure}{}

\preprint{1}

\title{Intermittency in Weak Magnetohydrodynamic Turbulence}

\author{Romain Meyrand}
\affiliation{Laboratoire de Physique des Plasmas, \'Ecole Polytechnique, F-91128 Palaiseau Cedex, France}
\email{romain.meyrand@lpp.polytechnique.fr}

\author{Khurom H. Kiyani}
\affiliation{Laboratoire de Physique des Plasmas, \'Ecole Polytechnique, F-91128 Palaiseau Cedex, France}
\affiliation{Centre for Fusion, Space and Astrophysics; Department of Physics, University of Warwick, Coventry, CV4 7AL, United Kingdom}

\author{S\'ebastien Galtier}
\affiliation{Laboratoire de Physique des Plasmas, \'Ecole Polytechnique, F-91128 Palaiseau Cedex, France}

\date{\today}

\begin{abstract}
Intermittency is investigated using decaying direct numerical simulations of incompressible weak   magnetohydrodynamic turbulence with a strong uniform magnetic field ${\bf b_0}$ and zero cross-helicity. At leading order, this regime is achieved via three-wave resonant interactions with the scattering of two of these waves on the third/slow mode for which $\kpa = 0$. When the interactions with the slow mode are artificially reduced the system exhibits an energy spectrum with $\kpe^{-3/2}$, whereas the expected exact solution with $\kpe^{-2}$ is recovered with the full nonlinear system. In the latter case, strong intermittency is found when the vector separation of structure functions is taken transverse to ${\bf b_0}$ -- at odds with classical weak turbulence where self-similarity is expected.  This surprising result, which is being reported here for the first time, may be explained by the influence of slow modes whose regime belongs to strong turbulence. We derive a new log--Poisson law, $\zeta_p = p/8 +1 -(1/4)^{p/2}$, which fits perfectly the data and highlights the dominant role of current sheets. 
\end{abstract}

\maketitle

One of the most striking features of strong hydrodynamic (HD) turbulence is the presence of both a complex chaotic spatial/temporal behavior and a remarkable degree of coherence. The small-scale correlations of turbulent motion are known to show significant deviations from Gaussian statistics usually expected in systems with a large number of degrees of freedom \citep{She88}. This phenomenon, known as intermittency, has been the subject of much research and controversy since Batchelor and Townsend's first experimental observation in 1949 \citep{Batchelor49}. It still challenges any tentative of rigorous analytical description from first principles (\ie from the Navier-Stokes equations). Intermittency can be measured with the probability density function (PDF) of the velocity differences between two points separated by a distance $\bell$. In the presence of intermittency, PDFs develop more and more stretched and fatter tails when $\bell$ decreases within the inertial range, showing the increasing probability of large extreme events. This non self-similarity of PDFs in HD reflects the fact that the energy dissipation of turbulent fluctuations is not space-filling \citep{K41} but concentrated in very intense vorticity filaments \citep{Douady91}. 

Recently, growing interest has been given to the study of intermittency in the weak turbulence (WT) regime \citep{naza}. WT is the study of the long-time statistical behavior of a sea of weakly nonlinear dispersive waves for which a natural asymptotic closure may be obtained \citep{naza}. The energy transfer between waves occurs mostly among resonant sets of waves and the resulting energy distribution, far from a thermodynamic equilibrium, is characterized by a wide power law spectrum that can be derived exactly \citep{ZLF}. WT is a very common natural phenomenon studied in, \eg capillary waves \citep{deike}, gravity waves \citep{Falcon}, superfluid helium and processes of Bose-Einstein condensation \citep{lvov03}, nonlinear optics \citep{Dyachenko}, rotating fluids \citep{Galtier2003} and space plasmas \citep{sun}. In particular, intermittency has been observed both experimentally and numerically, and is attributed to the presence of coherent structures \eg sea foam \citep{Newell92} or freak ocean waves \citep{Jansen03}. In these examples, intermittency is linked to the breakdown of the weak non-linearity assumption induced by the WT dynamics itself and therefore cannot be considered as an {\it intrinsic} property of this regime. In fact intermittency is at odds with classical WT theory because of the random phase approximation which allows the asymptotic closure and resultant derivation of the WT equations \citep{naza}. 

Weak magnetohydrodynamic (MHD) turbulence differs significantly from other cases because of the singular role played by slow modes for which $k_{\parallel}=0$; where $\textbf{k}$ is wavevector in Fourier space, and the subscript $\parallel$ indicates the component of $\textbf{k}$ parallel to the guide field ${\bf b_0}$. Since Alfv\'en waves have frequencies $\omega^{\pm}_{\textbf{k}}=\pm k_{\parallel}v_{A}$ (with $v_{A}$ the Alfv\'en speed) and only counter-propagating waves can interact, the resonance condition, $\omega^{+}_{\textbf{k}_{1}}+\omega^{-}_{\textbf{k}_{2}}=\omega^{\pm}_{\textbf{k}_{3}}$ and $\textbf{k}_{1}+\textbf{k}_{2}=\textbf{k}_{3}$, implies that at least one mode must have $k_{\parallel}=0$ \citep{Montgomery95}. This mode which acts as a catalyst for the non-linear interaction is not a wave but rather a kind of two-dimensional condensate with a characteristic time $\tau_{A} \sim 1/ \left(k_{\parallel}v_{A}\right)=+\infty$ and cannot be treated by WT. The standard way to overcome this complication has been to assume that the $k_{\parallel}$ spectrum of Alfv\'en waves is continuous across $k_{\parallel}=0$. Under this assumption the weak MHD theory was established by Galtier \textit{et al.} (2000) and a $\kpe^{-2}$ energy spectrum was predicted in the simplest case of zero cross-helicity with a direct cascade towards small-scales \citep{Galtier2000}. This prediction has been confirmed observationally \citep{Saur} and numerically \citep{Perez08} which indirectly might vindicate \textit{a posteriori} the continuity assumption.

In this Letter, we investigate weak MHD turbulence through three-dimensional high resolution direct numerical simulations. We use higher-order statistical tools to demonstrate the presence of intermittency in the cascade direction and show that this property can be understood via a log-Poisson law where the influence of slow modes, which belong to strong turbulence, is included. 

The incompressible MHD equations in the presence of a uniform magnetic field ${\bf b_0}$ read: 
\ba
\partial_t {\bf z^\pm} \mp b_0 \partial_{\parallel} {\bf z^\pm} + {\bf z^\mp} \cdot \nabla \, {\bf z^\pm} &=& 
- {\bf \nabla} P_* + \nu_3 \Delta^3 {\bf z^\pm} \, , \, \, \label{mhd1} \\
\nabla \cdot {\bf z^\pm} &=& 0 \label{mhd3} \, ,
\ea
where ${\bf z^\pm}={\bf v} \pm {\bf b}$ are the fluctuating Els\"asser fields, ${\bf v}$ the plasma flow velocity, ${\bf b}$ the normalized magnetic field (${\bf b} \to \sqrt{\mu_0 \rho_0} \, {\bf b}$, with $\rho_0$ a constant density and $\mu_0$ the magnetic permeability), $P_*$ the total (magnetic plus kinetic) pressure and $\nu_3$ the hyper-viscosity (a unit magnetic Prandtl number is taken). The MHD model offers a powerful description for large-scale astrophysical plasmas including solar/stellar winds, accretion flows around black holes and intracluster plasmas in clusters of galaxies \citep{biskamp}. Most often such plasmas are turbulent with an incompressible energetically dominant component and embedded in a large-scale magnetic field \citep{jltp}. 

Equations (\ref{mhd1})--(\ref{mhd3}) are computed using a pseudo-spectral solver (TURBO 
\citep{teaca}) with periodic boundary conditions in all three directions. Note that the nonlinear terms 
are partially de-aliased using a phase-shift method. Two situations will be considered: the full 
equations (case A) with $1536^2 \times 128$ collocation points (the lower resolution being in the $
{\bf b_0}$ direction where the cascade is negligible) and the case where the interactions with slow 
modes are artificially reduced (case B). In case B the reduction is obtained by imposing ${\hat 
{\bf v}}(\kpeb,\kpa=0)={\hat {\bf b}}(\kpeb,\kpa=0)=0$ at each time step (where $\hat{}$ denotes 
the Fourier transform). Note that it does not preclude totally the interactions between slow modes ($
\kpa=0$) and other wave modes ($\kpa>0$) due to the fact that the non-linear terms are 
computed in real space. Consequently, during the time advanced slow modes may receive some 
energy which eventually can lead to their interactions with the other waves modes. The initial state consists of magnetic 
and velocity field fluctuations with random phases such that the total cross-helicity is zero, and the 
kinetic and magnetic energies are equal to $1/2$ and localized at the largest scales of the system 
(mostly wave numbers $k\in [2,4]$ are initially excited). There is no external forcing and we fix $\nu_3 
= 4 \times 10^{-15}$ and $b_0=20$. Our analysis is systematically made at a time when the mean 
dissipation rate reaches its maximum.

\begin{figure}
\centering\subfigure[]{ \includegraphics[clip, scale=0.289,angle=-90]{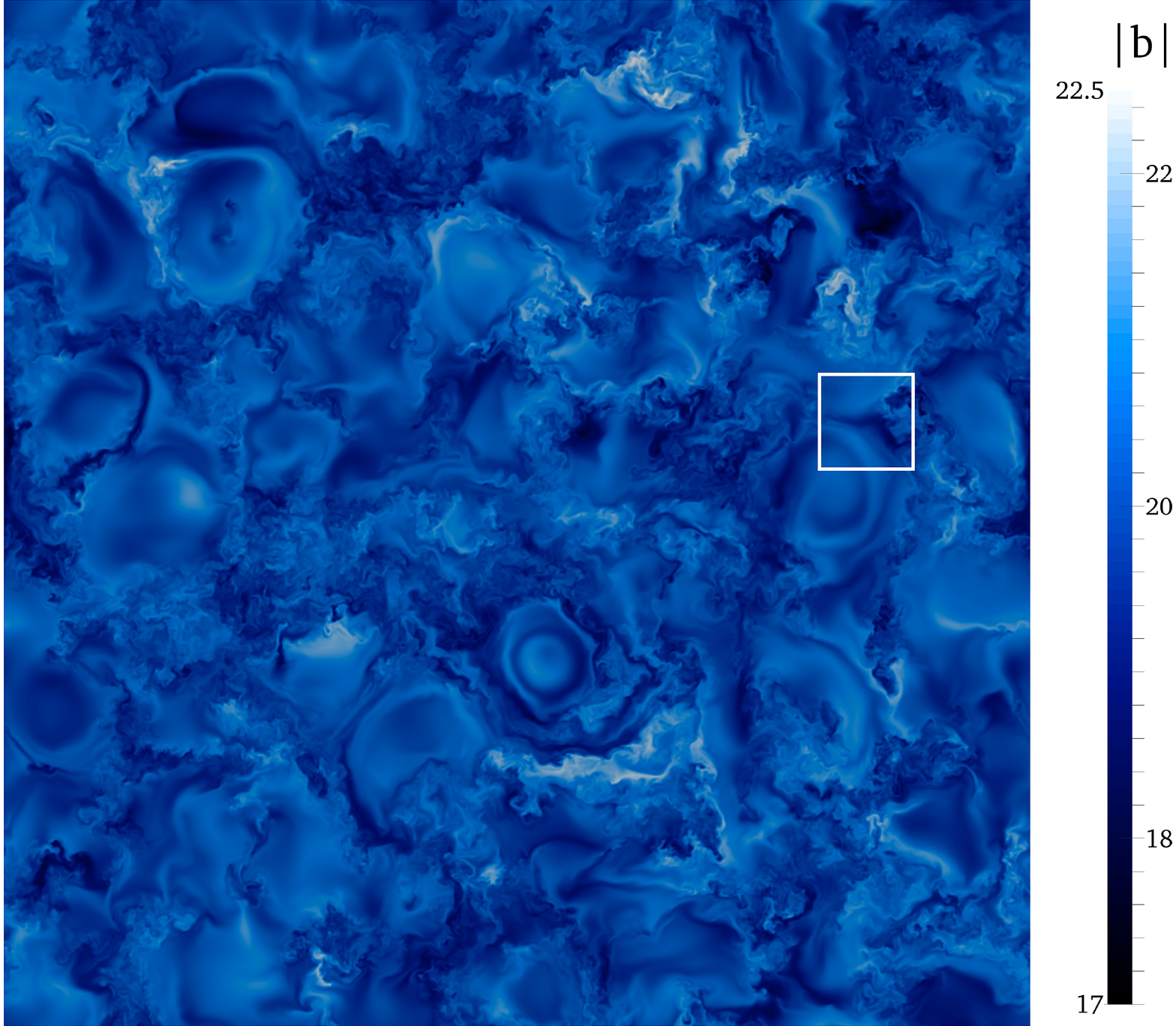}}
\subfigure[]{ \includegraphics[clip,scale=0.206]{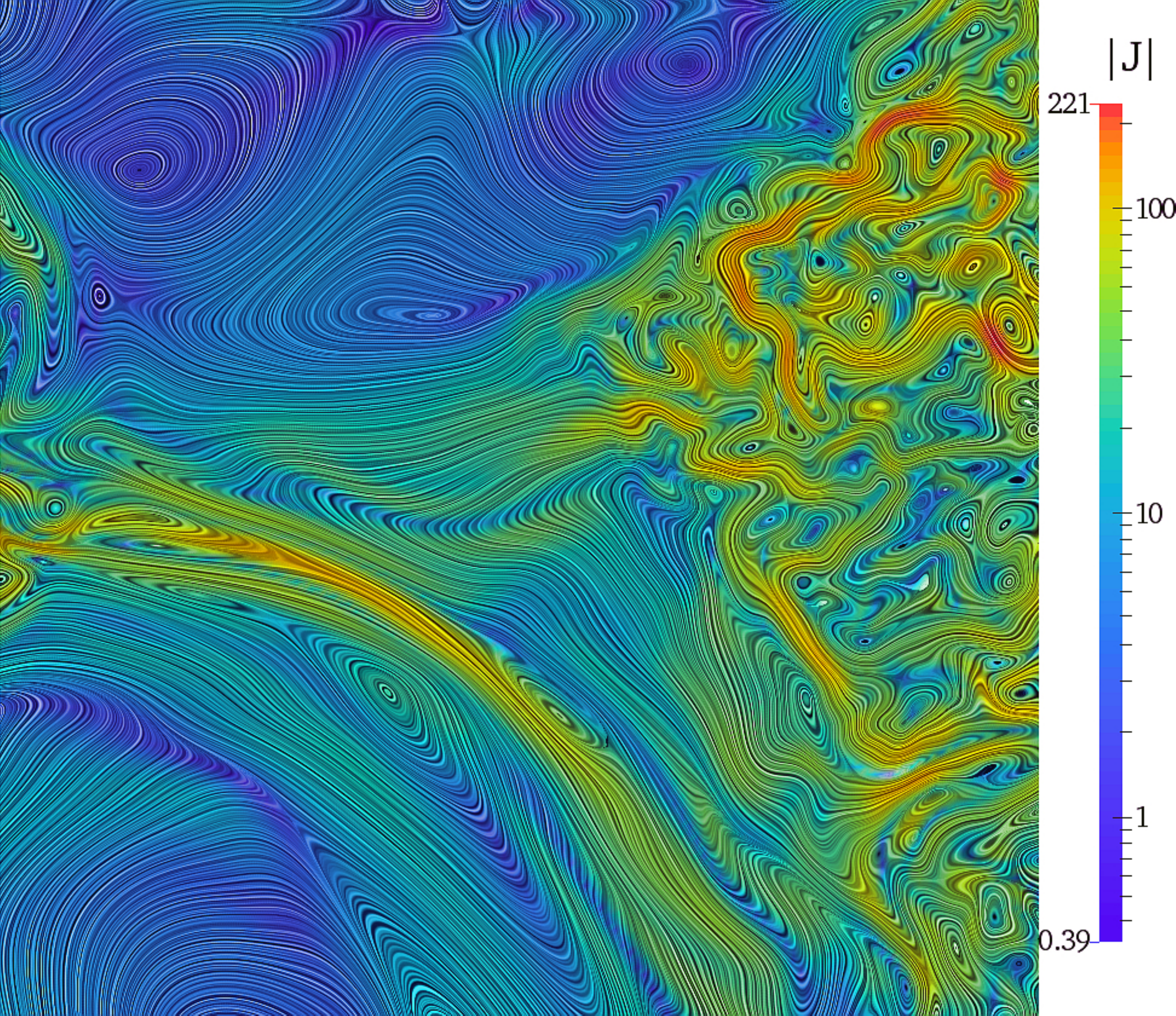}}
\caption{\textit{Top}: Snapshot of the magnetic field modulus (in linear scale) in a section perpendicular to ${\bf b_0}$ (case A). \textit{Bottom}: Close-up of the current density modulus (in logarithmic scale) corresponding to the marked (square) region on the top. The line integral convolution (LIC) technique \citep{LIC} reveals a hierarchy of current sheets as well as the formation of less intense filaments.}
\label{Fig1n}
\end{figure}
Fig. \ref{Fig1n} (top) shows a snapshot of the magnetic field modulus in a section perpendicular to ${\bf b_0}$ for case A. The large-scale coherent structures are mostly the signature of the initial condition whereas the incoherent small-scale structures are produced by the nonlinear dynamics and the direct energy cascade. It is believed that such patchy structures characterize the weak MHD turbulence regime. A close up of the current density modulus is also given (bottom). It reveals a hierarchy of current sheets which are the dominant dissipative structures. Interestingly, we also see the formation of less intense filaments.

We shall quantify the turbulence statistics by introducing the bidimensional axisymmetric Els\"asser energy spectra $E^\pm(\kpe,\kpa)$ which are linked to the Els\"asser energies ${\cal E}^\pm = \langle {\bf z^\pm}^2 \rangle / 2$ of the system ($\langle \, \rangle$ denotes an integration over the physical space) by the double integral ${\cal E}^\pm = \iint E^\pm(\kpe,\kpa) d\kpe d\kpa$. 
\begin{figure}
\centering
\includegraphics[scale=0.3]{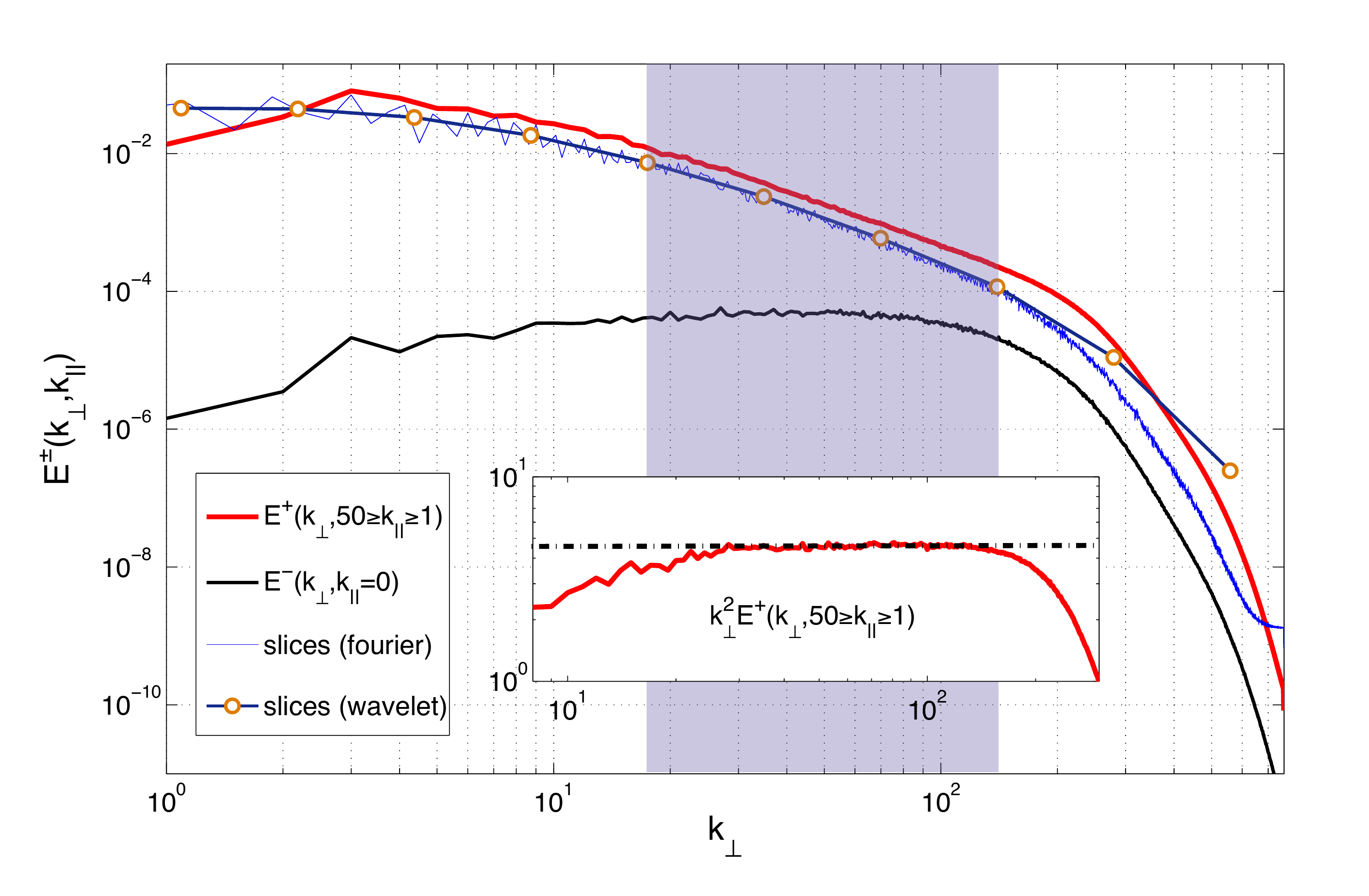}
\caption{Transverse wave number Els\"asser energy spectra, $\int_1^{50} E^+(\kpe,\kpa) d\kpa$ (red) and $E^-(\kpe,\kpa=0)$ (black) for case A. The former spectrum is also shown when it is compensated by $\kpe^2$ (inset). The wavelet spectrum from the sampled slices (see text) is also shown (blue). The shaded area corresponds to the scales for which the intermittency study is conducted. 
\label{Fig2n}}
\end{figure}
Fig. \ref{Fig2n} shows the results for case A. In order to improve the statistics the spectrum $E^+$ is plotted after an integration from $\kpa=1$ to $50$ (red). In this way we suppress slow modes contribution and limit the cumulative contribution of the dissipative parallel scales which can eventually alter the scaling law at the smallest scales. A spectrum compatible with the weak turbulence prediction in $\kpe^{-2}$ is clearly observed (see inset). Additionally, we over-plot the spectrum $E^-$ for the slow mode in order to show that it behaves very differently with a flat spectrum. The spectrum $E^-$ (or $E^+$ in the latter case) behaves similarly, as is expected for zero cross-helicity. 

\begin{figure}
\centering
\includegraphics[scale=0.47]{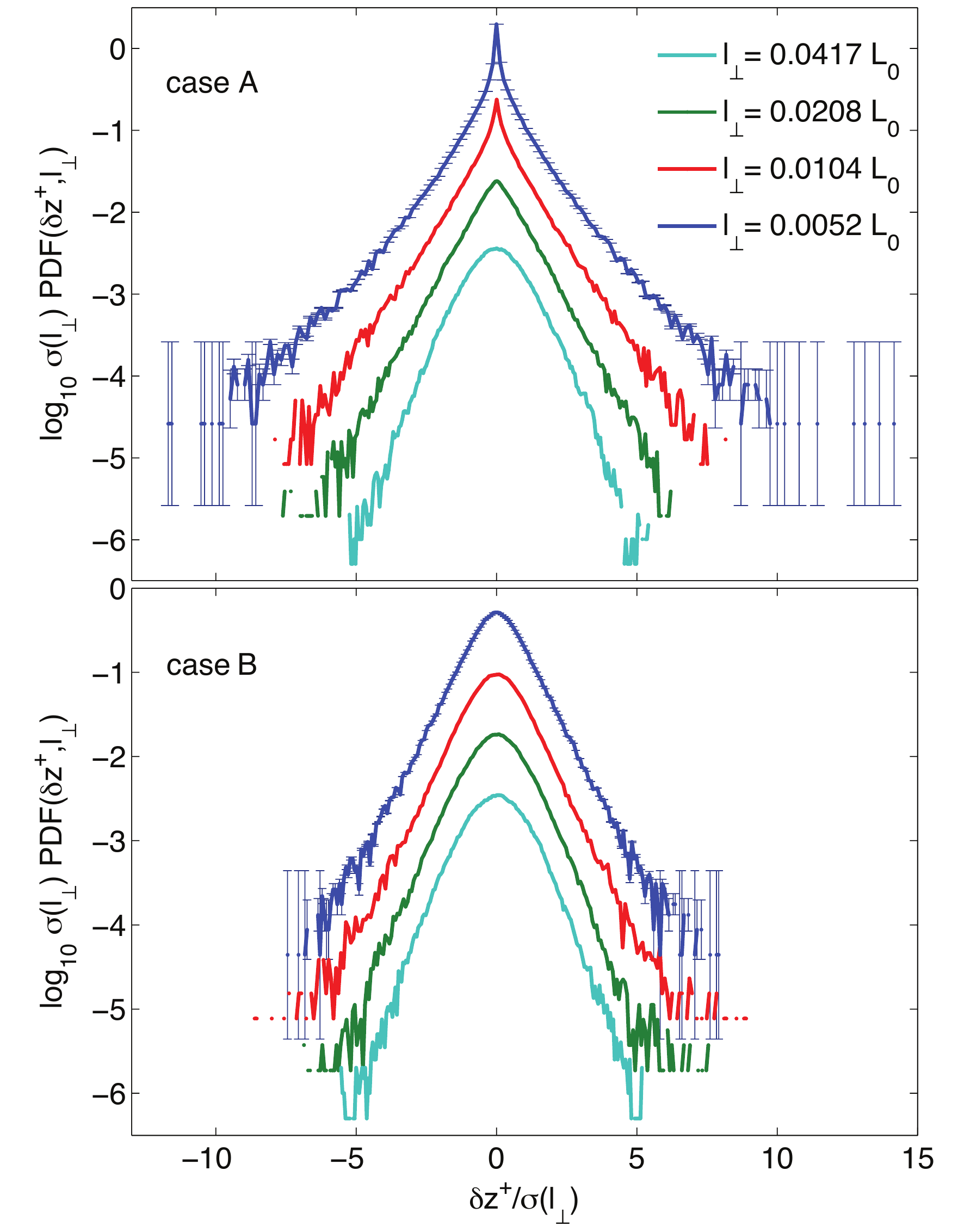}
\caption{PDFs of the Els\"asser field increments (given by wavelet coefficient) $\delta z^+$ for case A (top) and B (bottom), and for four distances $\ell_\perp$ ($L_0$ is the size of the numerical box) corresponding to the circles inside the shaded area in Fig. \ref{Fig2n}. The PDFs are shifted vertically for clarity. 
\label{Fig3n}}
\end{figure}

We use wavelet coefficients to define the Els\"asser field increments, denoted by $\delta z^\pm ({\bf x},\bell_\perp)$, between two points separated by a vector $\bell_\perp$ transverse to ${\bf b_0}$. For more details on the wavelet method used and the justifications for using it please refer to \cite{Kiyani2013}.
Intermittency can be investigated through the PDFs of these increments for differente distances $\ell_\perp$. (We do not report an intermittency analysis for vector separations along ${\bf b_0}$ because the corresponding range of scale is too narrow.) The results from this analysis are shown in Fig. \ref{Fig3n}. For case A (top), strong intermittency is revealed through the development of more extended and heavier tails at shorter distance $\ell_\perp$. The result is drastically different when the interaction with the slow mode is artificially reduced (case B, bottom) -- in this case intermittency is strongly reduced, with PDFs approaching closer to a Gaussian. Removing the interactions between  slow modes and other wave modes is equivalent to removing the resonant interactions which support the weak turbulence dynamics. Therefore, what we see in case B is mainly the result of the non-resonant triadic interactions. 

\begin{figure}
\centering
\includegraphics[trim=0.1cm 7.1cm 0.1cm 7.1cm, clip=true, scale=0.39]{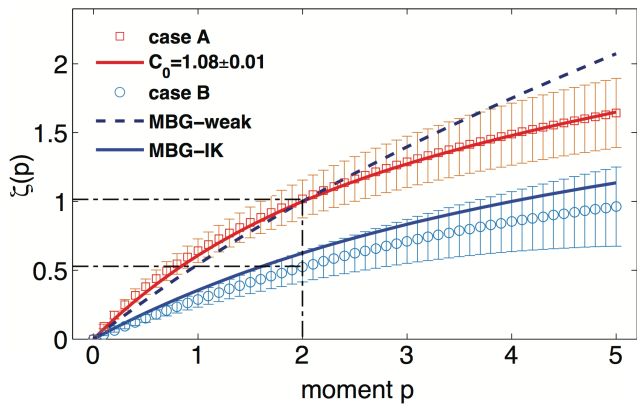}
\caption{ Scaling exponents $\zeta(p)$ for case A and B  with the log-Poisson model (Eq.\ref{weaklaw}) over-plotted for $C_0=1.08$. The models MBG-weak and MBG-IK are also display (see details in text) for comparison. The errors quoted correspond to one standard error of the parameter fits.
\label{Fig4n}}
\end{figure}

We further analyze intermittency through the symmetric structure functions:
\be
S_p = \left\langle \left( \delta z^+ \right)^{p/2} \right\rangle 
\left\langle \left( \delta z^- \right)^{p/2} \right\rangle \sim \ell_\perp^{\zeta(p)} \, , 
\ee
where $\zeta(p)$ are the scaling exponents that are to be measured in the inertial range (homogeneous axisymmetric turbulence is assumed). The study is conducted on several transverse planes and the fluctuations at different\sout{s} scales $\ell_{\perp}$ are calculated using the undecimated discrete wavelet transform detailed in \cite{Kiyani2013}; a 12-point wavelet was chosen to overcome the limitations of normal two-point structure functions (see \cite{Kiyani2013,Cho2009}). Utilizing the periodic boundary conditions of the simulation allows us to construct large contiguously sampled signals in each plane; the ensemble of fluctuations is then constructed from a union of all the fluctuations generated from the signal in each of the planes. This ensures that we have a large sample to form our statistics. Figure \ref{Fig4n} shows that weak turbulence (case A) is characterized by strong intermittency such that $\zeta(p)$ cannot be fitted with a trivial (linear) law, nor by the MHD log-Poisson model previously derived (\citep{muller}, Eq. (3)) when $g=2$ (hereafter MBG-weak)  or $g=4$ (hereafter MBG-IK) which correspond respectively to weak \citep{Galtier2000} and strong \citep{IK} turbulence. We also report the scaling exponents for case B which behaves differently as a result of the $k_\parallel=0$ modes being removed. Interestingly, for case B, the data follow the same curvature as the strong MHD turbulence model (MBG-IK) and, as can be seen from the value of $\zeta(2)$, are compatible with a $\kpe^{-3/2}$ spectrum. 

We want to build a model that fits the exact solution of weak turbulence (energy spectrum in $\kpe^{-2}$) which in physical space implies $\zeta(2) = 1$. Even if this power law relation between the Fourier and physical spaces is not mathematically exact, it is very well observed in Fig. \ref{Fig4n} and can be considered as an excellent constraint for the intermittency model. Following the original development \citep{she} we then define:
\be
S_p = C_p \langle \varepsilon^{p/2} \rangle \ell^{p/2}_\perp \, , 
\ee
where $C_p$ are some constants, $\langle \varepsilon \rangle $ is the mean dissipation rate of energy and $\langle \varepsilon^{p/2} \rangle \sim \ell_\perp^{\mu_{p/2}}$. The latter relation is the so-called refined similarity hypothesis \citep{K62}. The log--Poisson distribution for the dissipation leads to the general relation \cite{she,biskamp}:
\be
\mu_m = \mu(m) = -m\Delta + C_0 (1-\beta^m) \, , 
\ee
where $\Delta$ and $\beta$ are linked to the co-dimension $C_0$ of the dissipative structures such that $C_0 = \Delta / (1-\beta)$. We shall consider the co-dimension as a free parameter that will be estimated directly from the data. The system is closed by defining the value of $\Delta$ which is related to the dissipation of the most singular structures, such that $\ell_\perp^{-\Delta} \sim {E_\infty / \tau_\infty}$, where $E_\infty$ is the energy dissipated in these most singular structures and $\tau_\infty \sim \ell / v_\ell$ is the associated time-scale. $\Delta$ may be obtained by considering the following remarks. Weak MHD turbulence behaves very differently from isotropic MHD because in the former case the regime is driven at leading order by three-wave resonant interactions, with the scattering of two of these waves on the slow/third mode. (In fact, weak MHD turbulence is not applied in a thin layer around $\kpa=0$ \cite{Galtier2000} where strong turbulence is expected.) These slow modes are also important to characterize dissipative structures (see Fig. \ref{Fig1n}): these structures, which look like vorticity/current sheets, are strongly elongated along the parallel direction and are therefore mainly localized around the $\kpa = 0$ plane in Fourier space. If we assume that the dynamics of slow modes are similar to the dynamics of two-dimensional strong turbulence, then it seems appropriate to consider that the time-scale entering in the intermittency relation may be determined by \citep{biskamp} $v_\ell \sim \ell_\perp^{1/4}$, hence the value $\Delta=3/4$. Note that the importance of slow modes on intermittency was already emphasized in Fig. \ref{Fig3n} where the PDFs are closer to a Gaussian when the interactions with them are strongly reduced. With this all considered, we finally obtain the intermittency model for weak MHD turbulence:
\be
\zeta(p) = {p \over 8} + C_0 - C_0\left(1- {3 \over 4 C_0}\right)^{p/2} \, . 
\label{weaklaw}
\ee
This model is over plotted in Fig. \ref{Fig4n} (top, red line) for a fractal co-dimension $C_0 = 1.08$ which is the best value fitting the data (a non-linear least-squares regression is used). The model fits perfectly the numerical simulation, thus we may infer that weak MHD turbulence is mainly characterized by vorticity/current sheets. Note that the parameter $\beta$ in this model measures the degree of intermittency: non-intermittent turbulence corresponds to $\beta=1$ whereas the limit $\beta=0$ represents an extremely intermittent state in which the dissipation is concentrated in one singular structure. According to the value obtained here $\beta \simeq 1/4$ (with $C_0 \simeq 1$), we may conclude that weak MHD turbulence appear more intermittent than strong isotropic MHD turbulence for which $\beta=1/3$.

In conclusion, this work presents direct numerical simulations of weak MHD turbulence where intermittency is found and modeled with a log-Poisson law, and where the slow mode plays a central role via the dissipative structures.
This is an important observation, in fact the most important and key message of this Letter. It has profound implications for our interpretations of plasma turbulence observations and provides objective insights to the normally heated discussions on what constitutes turbulence in systems such as plasmas which host a rich variety of waves and instabilities and at the same time are inherently nonlinear. The results of our work, seem to suggest that the quintessential signature of turbulence in the form of intermittency, is not simply a property of strong or fully developed turbulence; but also of weakly nonlinear systems. This fact has previously been ignored due to the over-emphasis in the literature on spectra (second order statistics hide by construction 
the role played by phases) and the random phase approximation. Our simulations show that phase synchronization plays an important role, even in WT, whilst retaining some of the exact analytical results pertaining to spectra.

\paragraph*{Acknowledgements.}
The computing resources for this research were made available through the UKMHD Consortium facilities funded by STFC grant number ST/H008810/1. This work was granted access to the HPC resources of [CCRT/CINES/IDRIS] under the allocation 2012 [x2012046736] made by GENCI. R.M. acknowledges the financial support from the French National Research Agency (ANR) contract 10-JCJC-0403.

\end{document}